\documentclass[amsmath,longbibliography,preprint]{revtex4-2}  
\usepackage{bbm}
\usepackage{mathrsfs}
\usepackage{amsmath}
\usepackage{amsfonts}
\usepackage[colorlinks=true,citecolor=blue,anchorcolor=blue]{hyperref}
\usepackage{graphicx,epstopdf}
\usepackage{subfigure}
\usepackage{epsfig}
\usepackage{dcolumn}
\usepackage{bm}
\usepackage{color}
\usepackage{natbib}
\usepackage{amssymb}
\usepackage{xcolor}
\usepackage{braket}
\usepackage{soul}

\begin{document}

\title{Nonparaxiality-triggered Landau-Zener transition in topological photonic waveguides}

\author{An Xie$^{1}$}
\author{Shaodong Zhou$^{1}$}
\author{Kelei Xi$^{1}$}
\author{Li Ding$^{1}$}
\author{Yiming Pan$^{2}$}
\author{Yongguan Ke$^{3}$}\email{keyg@mail2.sysu.edu.cn}
\author{Huaiqiang Wang$^{4}$}\email{hqwang@nju.edu.cn}
\author{Songlin Zhuang$^{1}$}
\author{Qingqing Cheng$^{1,5}$}\email{qqcheng@usst.edu.cn}

\affiliation{$^{1}$Shanghai Key Laboratory of Modern Optical System,  School of Optical-Electrical and Computer Engineering, University of Shanghai for Science and Technology, Shanghai 200093, China}
\affiliation{$^{2}$ Physics Department and Solid State Institute, Technion, Haifa 32000, Israel}
\affiliation{$^{3}$Guangdong Provincial Key Laboratory of Quantum Metrology and Sensing $\&$ School of Physics and Astronomy, Sun Yat-Sen University (Zhuhai Campus), Zhuhai 519082, China}
\affiliation{$^{4}$ National Laboratory of Solid State Microstructures, School of Physics, Nanjing University, Nanjing 210093, China}
\affiliation{$^{5}$State Key Laboratory of Terahertz and Millimeter Waves, City University of Hong Kong, Hong Kong, China}

\maketitle
\textbf{Photonic lattices have been widely used for simulating quantum physics, owing to the similar evolutions of paraxial waves and quantum particles. However, nonparaxial wave propagations in photonic lattices break the paradigm of the quantum-optical analogy. Here, we reveal that nonparaxiality exerts stretched and compressed forces on the energy spectrum in the celebrated Aubry-André-Harper model. By exploring the mini-gaps induced by the finite size of the different effects of nonparaxiality, we experimentally present that the expansion of one band gap supports the adiabatic transfer of boundary states while Landau-Zener transition occurs at the narrowing of the other gap, whereas identical transport behaviors are expected for the two gaps under paraxial approximation. Our results not only serve as a foundation of future studies of dynamic state transfer but also inspire applications leveraging nonparaxial transitions as a new degree of freedom.}\\

\noindent{\bf \textcolor{red}{INTRODUCTION}}\\
Photonic lattices such as waveguide arrays provide a versatile platform for investigating fundamental physics \cite{rechtsman2013photonic,lu2014topological,maczewsky2017observation,mukherjee2017experimental,blanco2018topological,zilberberg2018photonic,ozawa2019topological,mukherjee2020observation,wang2020localization}. Over the past two decades, many intriguing phenomena have been demonstrated in evanescently coupled waveguide arrays, including Landau-Zener (LZ) transition \cite{shevchenko2010landau,sun2016finite}, topological end modes \cite{cheng2015topologically,blanco2016topological,Cheng2019observation,song2019breakup,umer2020counterpropagating}, Anderson localization \cite{schwartz2007transport,lahini2008anderson,levi2011disorder,segev2013anderson} in disordered lattices, $etc$. The underlying principle relies on the analogy between the paraxial Helmholtz equation for electromagnetic waves and the Schr\"{o}dinger equation describing quantum particles. However,  the paraxial approximation is not always strictly satisfied in waveguide systems, and in fact nonparaxial light is quite ubiquitous in natural photonic systems, such as spin-orbit interaction of nonparaxial light \cite{allen1992orbital, bliokh2010angular}, nonparaxial Airy beams \cite{siviloglou2007accelerating, siviloglou2007observation, zhang2011plasmonic, minovich2011generation, li2011plasmonic, cheng2021achromatic} and other nonparaxial accelerating beams \cite{zhang2012nonparaxial,kaminer2012nondiffracting,chremmos2013nonparaxial}, $etc$. Recently, nonparaxiality has attracted growing interests,  which has been shown to play an important role in third-harmonic generation \cite{penjweini2019nonlinear} and asymmetric topological pumping \cite{cheng2022asymmetric}. Nevertheless, the study of nonparaxial wave propagation and related phenomena is still in its infancy, and it still remains elusive how nonparaxiality can benefit the field of photonics. To this end, demonstrating the largely overlooked functionality of nonparaxility by  well-known fundamental physical phenomena would be highly desired.

As a typical fundamental dynamics, LZ transition, a transition between two states in a quantum system driven under a time-dependent Hamiltonian, is frequently encountered in different physical fields. Classical analogs of LZ transitions have also been investigated in platforms such as atomic-optical system \cite{niu1996atomic,wilkinson1997experimental,niu1998landau}, coupled cavities \cite{spreeuw1990classical,zheng2021photon} and other two-state systems \cite{bouwmeester1995observation,longhi2012quantum,ding2020observation}. 
Interestingly, LZ transition in topological transport of edge states has recently been demonstrated in driven acoustic cavity systems, where onsite term is implemented with the frequency in each waveguide cavity of varying height\cite{chen2021landau}. In addition, since adiabatic condition is usually required in topological photonics, such as the topological pumping in periodically-modulated lattice systems \cite{kraus2012topological,wang2013topological,ke2016Topological,nakajima2016topological,lohse2016thouless,zilberberg2018photonic,ma2018experimental,cerjan2020thouless,fedorova2020observation,jurgensen2021quantized}, the nonadiabatic LZ transition is expected to break the topological transport \cite{titum2016anomalous,privitera2018nonadiabatic}. It is thus highly appealing to study the LZ transition in nonparaxial topological photonic systems, where the interplay between topology, adiabaticity, and nonparaxiality could give rise to interesting phenomena.

In this work, we study the LZ transition in a spatially-modulated waveguide array designed according to the celebrated topological Aubry-André-Harper (AAH) model, which conducts microwave in a nonparaxial way. The two topological boundary states (TBSs) localized at opposite boundaries in each of the bulk gap act as an effective two-level quantum system. In our system, the two TBSs can be coupled by the finite-size effect, with an avoided crossing point and level repulsion between them. The time-dependent linear driving through the avoided-crossing points was realized by tuning the structural configurations of the microwave waveguide. It is found that when taking the paraxial approximation, the TBSs in the two bulk gaps have nearly equivalent mini-gaps, and exhibit identical behaviors, namely, LZ transition or adiabatic pumping between TBSs, depending on the energy gap and driving frequency. Intriguingly, when the realistic nonparaxial modifications are considered, one TBS mini-gap gets enlarged, while the other is reduced. Remarkably, LZ transition was observed around the significantly reduced mini-gap induced by nonparaxiality, which should be otherwise absent under the paraxial approximation. Our experimental results are nicely consistent with both numerical simulations and theoretical analysis of the exact Helmholtz wave equation. Our work uncovers
the previously underestimated importance of nonparaxiality, and may lead to an unexplored richness beyond the quantum optical analogies, especially in topological photonics.

\noindent{\bf \textcolor{red}{RESULTS}}\\
\noindent{\bf Photonic AAH model}\\
We start from the one-dimensional AAH model \cite{kraus2012topological,liu2015localization,ke2016Topological} with a periodic spatial modulation of the onsite potential, which is described by the tight-binding Hamiltonian

\begin{equation}
\label{equation1}
 H_{A}=\sum_{m=1}^{N}\beta_m \hat{c}_m^{\dagger}\hat{c}_m +\sum_{m=1}^{N-1}\big[\kappa\hat{c}_m^{\dagger}\hat{c}_{m+1}+\text { h.c.}\big].
\end{equation}
Here, $m$ labels the lattice site with a total number of $N$, $\hat{c}_{m}^{\dagger}(\hat{c}_m)$ creates (annihilates) a particle at the $m$th site, $\kappa$ is the nearest-neighbour(NN) hopping coefficient, and $\beta_m=\beta_{0}+\Delta \beta\cos(2\pi bm+\phi)$ represents the spatially-modulated potential of the $m$th site. In the modulation term,  $\Delta\beta$ is the modulation amplitude, $b$ controls the periodicity, and without loss of generality it is set as $b=1/3$ throughout the paper, resulting in a supercell structure with three sublattices. $\phi\in [0,2\pi]$ is the modulation phase, which plays a similar role as a momentum variable.

\begin{figure}[htp!]
	\includegraphics[clip,angle=0,width=14cm]{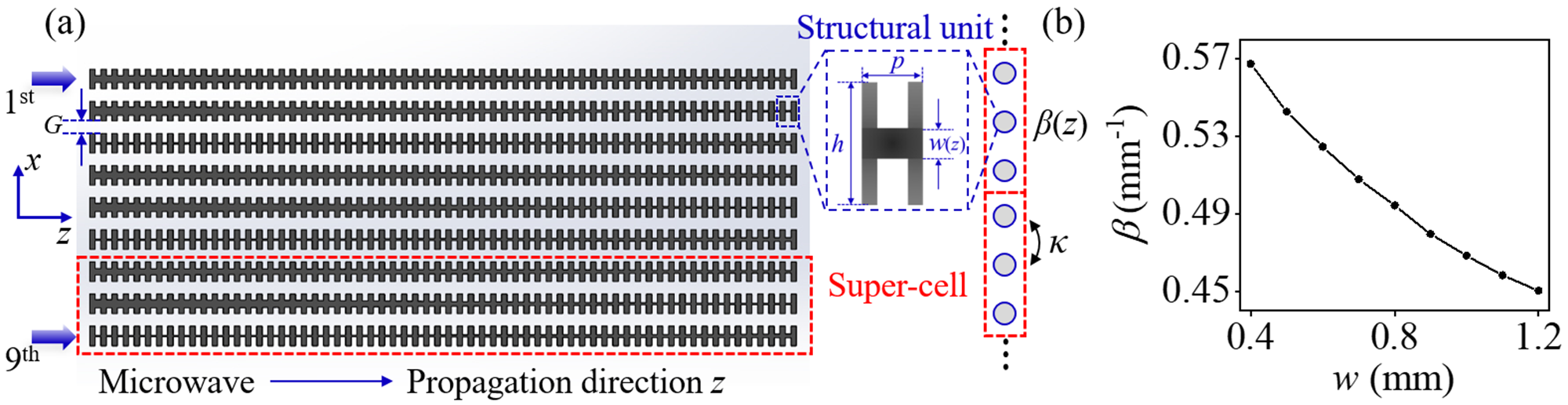}
	\caption{\textbf{Schematic diagram of a finite Aubry-André-Harper (AAH) lattice realized by well-designed periodically-modulated ultrathin metallic waveguides.} 
	(a) The constant nearest-neighboring (NN) hopping comes from NN coupling between equally-spaced waveguides, while the modulation of onsite potential (propagation constant) can be realized by changing the girder width $w$ of “H-shape” unit of each waveguide. 
	(b) Propagation constant $\beta$ as a function of girder width $w$.}
	\label{fig1}
\end{figure}

To realize a photonic counterpart of the above AAH model, we have experimentally fabricated  a well-designed ultrathin metallic waveguides composed of the ``H-shape" structural unit \cite{ma2014broadband,Cheng2018flexibly}, as schematically shown in Fig.~\ref{fig1}(a), where the $x$-direction denotes the spatial dimension and the propagation direction $z$ acts as the synthetic time dimension. Through the coupled-mode theory, the constant NN hopping $\kappa$ can be simulated by the coupling between equally-spaced NN waveguides in the whole array, with details of spacings and hoppings provided in Fig. S1. For the later realization of LZ transition, a time-dependent modulation of phase along the propagation direction is introduced as $\phi=\phi_{0}+\Omega z$, where $\phi_0$ represents the initial phase and the ``frequency'' $\Omega$ is defined as the ratio between the total change of the phase $\Delta\phi$ and the waveguide length $L$. The desired instantaneous onsite potential profile can then be obtained by modulating the propagation constant of each waveguide along both the spatial dimension $x$ and propagation direction $z$, which is achieved by changing the girder width $w$ of the ``H-shape" unit, as shown in Fig.~\ref{fig1}(b).

\begin{figure}[htp!]
	\includegraphics[clip,angle=0,width=14cm]{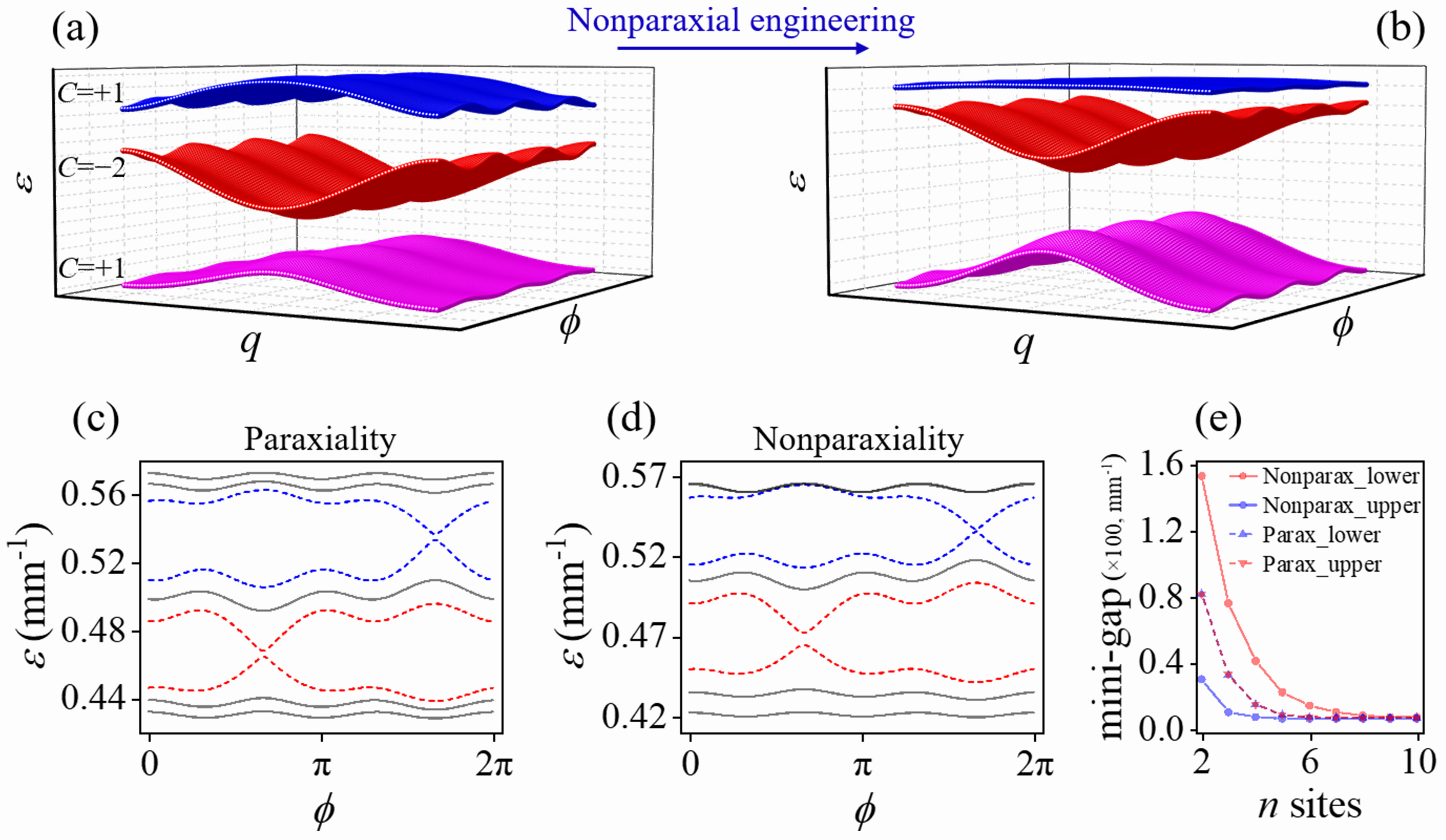}
	\caption{\textbf{The band structures in the two-dimensional momentum space ($q$, $\phi$) under (a) paraxial approximation and (b) nonparaxial modification.} The calculated eigenvalue as a function of modulation phase $\phi$ under (c) paraxial approximation and (d) nonparaxial correction.  The parameters are $N=9$, $\beta_{0}=0.501~\mathrm{mm}^{-1}$ ($w_{0}=0.8~mm$), $\kappa=0.0316 ~\mathrm{mm}^{-1}$($G$ = 1.8 mm),$\Delta \beta=0.0493~\mathrm{mm}^{-1}$ ($\Delta w=0.35 ~mm$) and $\kappa_{\mathrm{NNN}}=-0.00632~\mathrm{mm}^{-1}$ ($2kn_0\sim0.158$). 
    (e) The lower and upper gaps as a function of the number of lattice sites.}
    \label{fig2}
\end{figure}

By carrying out the Fourier transform $\hat{c}_{m}=(1/\sqrt{N})\sum_{q}e^{iqm}\hat{c}_{q}$ under periodic boundary conditions and treating $\phi$ as a momentum dimension,  the band structure in the two-dimensional (2D) $(q,\phi)$ momentum space can be obtained, as shown in Fig.~\ref{fig2}(a), which possesses three bands due to the three-sublattice structure~\cite{gomez2013floquet,atala2013direct}.  The topological property of each band can be characterized by the Chern number \cite{thouless1982quantized,niu1985quantized,kohmoto1985topological,ma2019topological} $(C)$ derived from the integral of the Berry curvature in the 2D $(q,\phi)$ plane. The Chern numbers for the three bands are calculated as $C=(1,-2,1)$, guaranteeing the emergence of chiral edge states [henceforth referred to as TBSs] traversing both bulk gaps when choosing open boundary conditions with finite waveguides along the $x$-direction. The TBSs are localized at the boundary waveguides, exhibiting a nearly exponential decay into the bulk. Consequently, for sufficiently large lattices, the 
TBSs from opposite boundaries within the same gap have negligible spatial overlap and coupling, thus forming a gapless crossing in each gap (see Fig. S2 with $N=45$ lattice sites). However, when reducing the lattice size, the spatial overlap between the TBSs gradually increases, and becomes non-negligible for small lattice sizes.  This will cause considerable coupling and level repulsion with a mini-gap between the two TBSs. 

Interestingly, when treating the TBSs in each bulk gap as a two-level system, the modulation along the propagation direction practically acts as a time-dependent driving of the two-level system \cite{llorente1992tunneling,barnes2012analytically}.  When the modulation frequency $\Omega$ is comparable to the mini-gap between the two TBSs, LZ transition may happen between them. In the following, we will go beyond the coupled-mode theory, and consider how the nonparaxiality of microwave modifies the energy spectrum and affects the LZ transition.\\

\noindent{\bf Nonparaxial modifications}\\

In the derivation of the above photonic AAH model from the coupled-mode theory, paraxial approximation has been assumed, which requires the condition of $\left|\partial^{2} \psi / \partial z^{2}\right| \ll 2 k|\partial \psi / \partial z|$ with the wave vector $k$. Specially, since our waveguide system conducts microwaves with much smaller wave vectors, the above condition will not be fully satisfied, and thus we need to go beyond the coupled-mode theory and turn back to the original Helmholtz equation for the waveguide system,

\begin{equation}
\label{equation2}
i \frac{\partial \psi}{\partial z}-\frac{1}{2 k n_{0}} \frac{\partial^{2} \psi}{\partial z^{2}}=H \psi. 
\end{equation}
Here, $H=\frac{1}{2kn_{0}}\nabla_{x}^{2}+\frac{k}{2}(\frac{n^{2}-n_{0}^{2}}{n_{0}})$ is the Helmholtz-Hamiltonian operator which can be replaced by the above structurally modeled tight-binding AAH Hamiltonian $H_{A}$, $n_{0}$ is the reference refractive index, and the wave vector is given as $k=\omega/c$ for monochronic electromagnetic wave $E=\Re\left\{\psi e^{i \omega t-i k n_{0} z}\right\}$. To achieve an intuitive understanding of the nonparaxial effect, we can rewrite Eq.~\eqref{equation2} in an effective Schrödinger-type form that absorbs  the second derivative of $z$ in a self-consistent way~\cite{cheng2022asymmetric}:

\begin{equation}
\label{equation3}
i \frac{\partial \psi}{\partial z}=\frac{H_{A}}{1+i \frac{1}{2 k n_{0} \partial z}} \psi=H_{\mathrm{eff}} \psi,
\end{equation}
where the effective Hamiltonian under Pad\'e approximation is approximately given by

\begin{equation}
\label{equation4}
H_{\mathrm{eff}}^{(1,1)} \approx H_{A}-\frac{1}{2 k n_{0}} H_{A}^{2}.
\end{equation}
Interestingly, apart from the paraxial Hamiltonian $H_A$, the consideration of nonparaxiality leads to an additional term proportional to $H_A^2$ (The lower $(m,n)$-order Padé approximations are listed in Table S2). Based on the previous work \cite{cheng2022asymmetric}, we could conclude that the nonparaxial term $-\frac{1}{2 k n_{0}} H_{A}^{2}$  contributes a negative next-nearest-neighboring (NNN) coupling, accordingly,  $\kappa_{\mathrm{NNN}}\approx-\kappa^{2}/\left(2 k n_{0}\right)$. For our microwave system with a very small wave vector $k$, such a negative NNN couplings cannot be neglected and reshapes the energy spectrum. In contrast, for larger wave vector (e.g., the optical region), the negative NNN couplings can be ignored, and the effective Hamiltonian is reduced to the paraxial one.

Next, we study how the nonparaxial term modifies the band structure. Thanks to the commutation relation $[H_{\textrm{eff}}^{(1,1)},H_A]=0$, the eigenstate $|\phi\rangle$ of $H_A$ with eigenvalue $\varepsilon$ is also an eigenstate of $H_{\textrm{eff}}$, but the effective energy is shifted to $\varepsilon-\frac{1}{2 k n_{0}} \varepsilon^{2}$. This modification results in a deformation of the original band structure, as shown in Fig.~\ref{fig2}(b). Since the modification is small, the order of the original bands remains unchanged, and the two band gaps are preserved. Moreover, the Chern number for each band also keeps invariant because of the same eigenstates as the original one, thus preserving the bulk topology as well as TBSs. However, it should be noted that for the experimentally relevant parameter regime with a small but non-negligible fitting parameter $1/(2kn_0)$, as can be seen in Fig. 2(b), the first (third) band obviously gets compressed (stretched) with significantly enlarged (reduced) band dispersion and band width.

\noindent{\bf LZ transition between TBSs}

The realization of LZ transition relies on two conditions, one is a two-level system with an avoided energy crossing determined by a control parameter, and the other is a time-dependent driving to sweep the parameter across the avoided crossing point, where the changing rate of the energy should be comparable to the minimum gap at the avoided crossing. Now we will first demonstrate that the first conditions can be satisfied in our photonic AAH model by using the two TBSs in each bulk band gap as the required two-level system, and then present our experimental results to illustrate how the nonparaxiality change the second condition..

For the condition of avoided level crossing, we resort to the finite-size effect, where the two TBSs become coupled and repulsed to each other due to considerable overlap of their wavefunctions. As a concrete example, we consider a small system with $N=9$ lattices (three super cells) and calculate the energy spectra as a function of modulation phase $\phi$ for both the paraxial Hamiltonian $H_A$ (without NNN hopping terms)
and nonparaxial effective Hamiltonian $H_{\textrm{eff}}$ 
(with NNN hopping terms $\kappa_{\mathrm{NNN}}=-0.00632~\mathrm{mm}^{-1}$)
, as shown in Figs.~\ref{fig2}(c) and ~\ref{fig2}(d), respectively. The other parameters are chosen as $\beta_{0}=0.501~\mathrm{mm}^{-1}$($w_{0}=0.8~mm$), $\kappa=0.0316 \mathrm{~mm}^{-1}$ , and $\Delta \beta=0.0493~\mathrm{mm}^{-1}$ ($\Delta w=0.35~mm$). 

In the paraxial case, as shown in Fig. 2(c), two nearly equivalent mini-gaps between two TBSs are opened at the crossing points around $\phi=2\pi/3$ and $\phi=5\pi/3$, respectively, which results from almost the same energy dispersion and hence the group velocity of the first and third bulk bands. 
In the nonparaxial case, however, the upper (lower) mini-gap is reduced (enlarged). 
The different mini-gap can be understood in the following way.
Assume the energy spacing around $\varepsilon$ is $\delta \varepsilon$ for the paraxial case.
According to the effective energy $\varepsilon-\frac{1}{2 k n_{0}} \varepsilon^{2}$, the energy spacing in the nonparaxial case is given by $[1-1/(kn_0)\varepsilon]\delta \varepsilon$, resulting in that the higher energy has smaller energy spacing.
In Fig.~\ref{fig2}(e), we show the lower and upper mini-gaps of TBSs at the avoided crossing points as a function of system size. As the system size increases,  the general decrease of two mini-gaps can be explained by the weaker and weaker higher-order couplings between TBSs. It means that the nonparaxial effects on LZ transition are more prominent in small-size system.

\begin{figure}[htp!]
	\includegraphics[clip,angle=0,width=14cm]{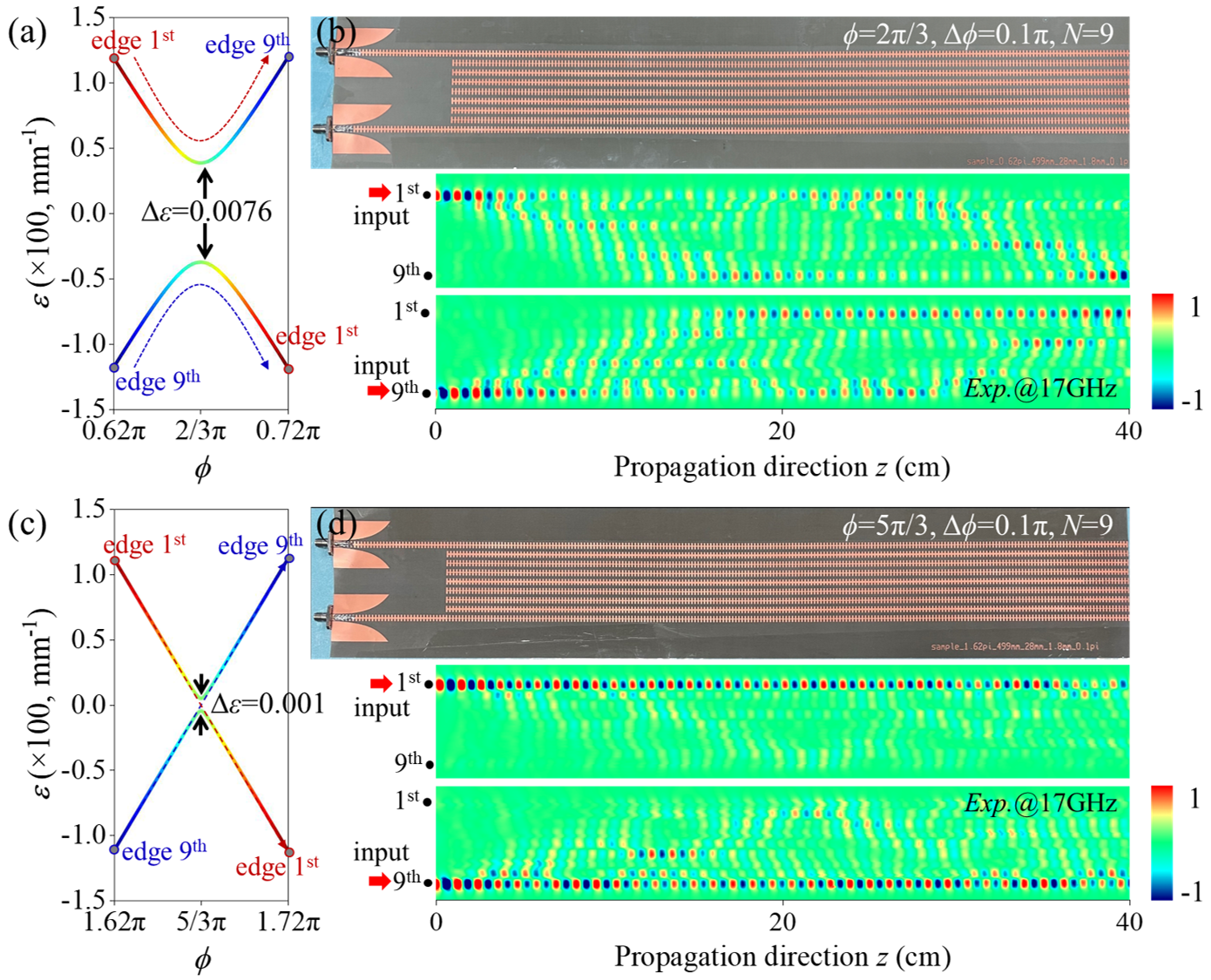}
	\caption{\textbf{Experimental results of topological pumping and LZ transition in microwave waveguide array.} (a) and (c) Energy spectrum around $\phi=2\pi/3$ and $\phi=5\pi/3$, respectively. (b) and (d) Schematic diagram of waveguide array and the observation of pumping process around $\phi=2\pi/3$ and $\phi=5\pi/3$, respectively. The other parameters are chosen as $N=9$, $\Delta\phi=0.1\pi$, $G=1.8$ mm, $L=40$ cm.}\label{fig3}
\end{figure}

The other condition of a time-varying control parameter is engineered by time-dependent modulation phase $\phi$ in our system. It should be emphasized that $\phi$ can be easily tuned along the synthetic time direction $z$ of the waveguide system, thus providing us with high flexibility in engineering LZ transition. In the vicinity of the avoided crossing of the TBSs, LZ transition can be modeled by a two-level effective Hamiltonian \cite{zhu1995theory,nakamura2012nonadiabatic},

\begin{equation}
\label{equation5}
H_{LZ}(\delta \phi)=\left(\begin{array}{cc}
\alpha \delta \phi & \frac{\Delta \varepsilon}{2} \\
\frac{\Delta \varepsilon}{2} & -\alpha \delta \phi
\end{array}\right).
\end{equation}
Here,  the coupling term $\Delta \varepsilon$ can be obtained by the minimum gap size at $\delta\phi=0$, and $\alpha$ is a fitting parameter characterizing the slope of the crossing. The basis of $H_{LZ}$ are $|\psi_{T}\rangle$ and $|\psi_{B}\rangle$ i.e., the TBSs localized at the top and bottom boundaries, respectively. We focus on the experimentally-relevant nonparaxial case with $N=9$, and plot the eigenvalues of $H_{LZ}$ around the two avoided crossings at $\phi=2 \pi/ 3$ and $\phi=5\pi/ 3$ (solid curves) in Figs.~\ref{fig3}(a) and ~\ref{fig3}(c), respectively. The eigenstates turn out to be hybridized states of $|\psi_{T}\rangle$ and $|\psi_{B}\rangle$ as a result of the coupling. The degree of hybridization is characterized by different colors, where the red and blue colors represent the two unhybridized limits of wavefunctions $|\psi_{T}\rangle$ and $|\psi_{B}\rangle$ localized at top and bottom boundaries, respectively. 

Intriguingly, by taking advantage of the significantly enlarged (reduced) gap at $\phi=2  \pi/3$  ($\phi=5  \pi/3$) induced from the nonparaxial effect, we now show that both adiabatic transport behavior and nonadiabatic LZ transition can be realized in our system under the same modulation frequency $\Omega$. This can be achieved by choosing $\Omega$ in the range smaller than the gap at $\phi=5\pi / 3 $ but larger than the gap at $\phi=5\pi / 3$. In our system, we consider a linear growth of modulation phase $\Delta\phi=0.1\pi$ within the whole waveguide length of $Z_{\textrm{max}}=400$ mm. The changing rate $\Omega$ is given by ${\Delta \phi}/{Z_{\max }}$. Such a choice makes it possible that the state evolves adiabatically along the eigenstates around $\phi=2\pi/3$ while it undergoes LZ transition around $\phi=5\pi/3$. In the former case of adiabatic transport, we assume that the initial states at $\phi_0=0.62 \pi$ are sufficiently far away from the avoided crossing of $|\psi_{T}\rangle$ and $|\psi_{B}\rangle$ at $\phi=2 \pi/3$. As shown in Fig.~\ref{fig3}(a), the initial top boundary state (red) could evolve adiabatically with increasing $\phi$ into the final state (blue) at the bottom boundary, and vice versa. In the latter case of LZ transition, we also consider the initial states of $|\psi_{T}\rangle$ and $|\psi_{B}\rangle$ at $\phi_0=1.62 \pi$ away from the avoided crossing at $\phi=5\pi/3$. As can be seen in Fig.~\ref{fig3}(c), when approaching the avoided crossing with increasing $\phi$, the top boundary state (red) in the upper energy level tunnels to the top boundary state in the lower level, and so does the bottom boundary state (blue). This means that the LZ transition renders each boundary state localized at where they start with negligible transfer to the other boundary.


To make it clearer, considering the initial microwave injected into the top port ($\psi(0)\approx|\psi_{T}\rangle$), we can directly apply the formula of LZ transition \cite{vutha2010simple,ke2015bloch} and obtain that the intensity distribution of the final state is proportional to $|\psi_T|^2=e^{\frac{-\pi}{\alpha} V}$ at the top port and $|\psi_B|^2=1-e^{\frac{-\pi}{\alpha} V}$ at the bottom port, where $V=\frac{1}{4}\times{(\Delta \varepsilon)^{2}}/{\left(\Delta \phi / Z_{\max }\right)}$. When the pumping center is located at $\phi_{0}=2\pi/3$, we can get $\alpha=0.04 \mathrm{~mm}^{-1}$ and $\frac{\pi}{\alpha} V \gg 0$, leading to $|\psi_B|^2 \approx 1$. In this case, LZ transition is negligible and the top-boundary state can be successfully transferred to the bottom-boundary state. Similarly, when the pumping center is located at $\phi_{0}=5\pi/3$, we can also derive $\alpha=0.03 \mathrm{~mm}^{-1}$ and $\frac{\pi}{\alpha} V \approx 0 $, resulting in $|\psi_T|^2 \approx 1$. It means that the almost complete LZ transition happens and the final state is localized at the top port (see SM for details of electric field propagation in our platform simulated by CST).

In experiments, the samples of microwave waveguides have the same parameters (N=9, $\Delta \phi=0.1 \pi$, G=1.8 mm, and L=400 mm) as the above discussion; see Fig.~\ref{fig3}(b) for $\phi_{0}=0.62\pi$ and Fig.~\ref{fig3}(d) for $\phi_{0}=1.62\pi$. After injecting the electric field into the samples, we detect the propagation of electric fields on the microwave near field platform (see the SM for the results of experiment and simulation in the case of  N=6). As expected, in the case of $\phi_{0}=0.62\pi$, we observe that the electric field injected from the $1^{\mathrm{st}}$(or $9^{\mathrm{th}}$) boundary waveguide gradually transfers across the bulk to the $9^{\mathrm{th}}$ (or $1^{\mathrm{st}}$) boundary waveguide. Note that the two pumping processes from the top-port excitation and bottom-port excitation are symmetric, as a result of the symmetric energy structure around the avoided-crossing points in Fig.~\ref{fig3}(a). The electric fields around the avoided crossing point are dominated in both the top and bottom ports, which can be viewed as a effect of beam splitter. In the case of $\phi_{0}=1.62\pi$, the electric field injected from the $1^{\mathrm{st}}$ (or $9^{\mathrm{th}}$) waveguide keeps rather stable during the pumping process, so that the electric field is still concentrated on the initial waveguide $1^{\mathrm{st}}$ (or $9^{\mathrm{th}}$) at the output end. Our results already demonstrate the successful transfer of TBS and the localization of isolated TBS, which is based on the nonparaxial condition in the LZ model realized in the microwave system.

\begin{figure}[htp!]
	\includegraphics[clip,angle=0,width=10cm]{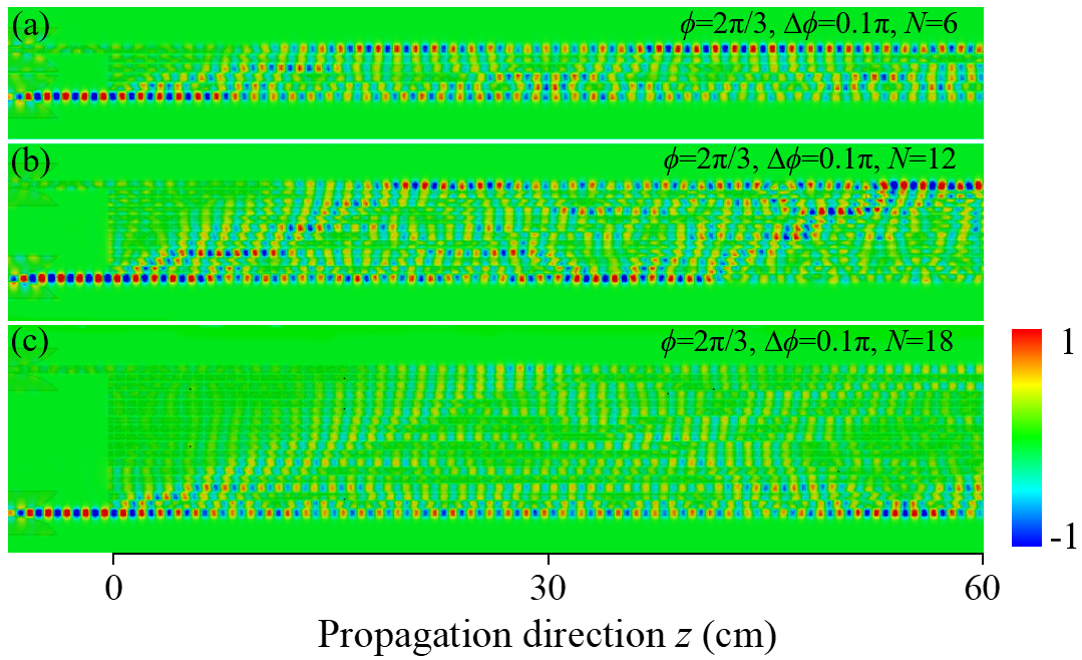}
	\caption{\textbf{CST simulation of pumping processes with different sites (a) $N=6$ and (b) $N=12$, and (c) $N=18$.} The injected light propagates along the input boundary waveguide without scattering. The other parameters are chosen as $L=60$ cm, $\phi=2\pi/3$, $\Delta\phi=0.1\pi$, Spacing $G=1.8$ mm between adjacent waveguides.\label{fig4}}
\end{figure}

In addition, it is worth emphasizing that the nonparaxial effects (differences in energy gap and field propagation) are significant with a small number of waveguides. We already show the gap around the avoided-crossing point would get smaller as the system size increases. For a fixed modulation frequency, we may observe that the field propagation changes from adiabatic evolution to LZ transition with increasing waveguide number N.  To confirm our argument, the CST simulations have been shown in Fig.~\ref{fig4} for the three cases: $N=6, N=12$, and $N=18$. From the results of simulations, in the cases of $N=6$ and $N=12$, the adiabatic tunneling from the right to left boundaries could be observed, but the coupling between boundary states is blocked as the N increases to 18, which demonstrates the negligible gap and complete LZ transition.

\noindent{\bf \textcolor{red}{DISCUSSION}}\\
In summary, we have theoretically predicted and experimentally demonstrated nonparaxiality-triggered LZ transition in microwave waveguide arrays with both spacial and temporal modulations. Both the adiabatic pumping and LZ transition can be analyzed in a simplified coupled two-level LZ model of TBSs. Nonparaxiality of microwaves plays a crucial role in modifying spectrum and the adiabatic condition, that is, one gap is enlarged and supports the adiabatic transfer of boundary states while the other gap is narrowed to make LZ transition, whereas identical transport behaviors are expected for the two gaps under paraxial approximation. Our approach of microwave nonparaxial engineering can be extended to higher dimensions and can benefit other physical systems, such as mechanical vibrations, elastic waves, electrical circuitries, and thermal transfers. We also expect our work to open up more directions on topological waves with the new potentials for topological devices. 

\noindent {\bf \textcolor{red}{SUPPLEMENTARY MATERIALS}}\\
Supplementary material for this article is available at XXX

\noindent {\bf \textcolor{red}{REFERENCES AND NOTES}}

\bibliography{reference}

\noindent {\bf Acknowledgement:}
The authors are thankful to Prof. Yuri S. Kivshar and Prof. Chaohong Lee for helpful discussions.

\end{document}